\documentclass[aps,pra,twocolumn,nofootinbib]{revtex4}
\usepackage{amsmath, amsthm, amscd, amssymb}
\usepackage{bm}
\usepackage{bbm}
\usepackage{epsfig}

\begin{document}

\title{Comment on ``Tests of general relativity with GW170817"}
\author{Anatoly A. Svidzinsky$^1$ and Robert C. Hilborn$^2$}
\affiliation{$^1$Department of Physics \& Astronomy, Texas A\&M University, College
Station, TX 77843 \\
$^2$American Association of Physics Teachers, One Physics Ellipse, College
Park, MD 20740 }
\date{\today }

\begin{abstract}
In a recent paper \textquotedblleft Tests of general relativity with
GW170817\textquotedblright\ (arXiv:1811.00364 [gr-qc]) the authors claimed
overwhelming evidence in favor of tensor gravitational wave (GW)
polarization over vector by analyzing GW signals measured by the LIGO-Virgo
network. Here we show that the measured LIGO-Livingston signal is substantially
reduced at certain frequency intervals which can be attributed to noise
filtering. We found that if these regions are excluded from the analysis
then data are consistent with vector GW polarization and not with tensor. We
show that if the signal accumulation method is applied over the entire
detector bandwidth, including the regions in which the signal is depleted by
noise subtraction, the result underestimates the LIGO-Livingston signal
amplitude. That smaller amplitude then leads to an erroneous conclusion that
favors tensor polarization over vector polarization for the GW.
\end{abstract}

\maketitle

In a recent preprint \cite{Max18} the authors reported results of the
gravitational wave (GW) polarization test with GW170817 performed using a
Bayesian analysis of the signal properties with the three LIGO-Virgo
interferometer outputs. The authors found overwhelming evidence in favor of
pure tensor polarization over pure vector with an exponentially large Bayes
factor. This result is opposite to the conclusion of our analysis based on a
direct comparison of the GW signals measured by the three detectors \cite%
{Svid18a}. Namely, we found that the measured signal ratios are inconsistent
with the predictions of general relativity, but consistent with the recently
proposed vector theory of gravity \cite{Svid17,Svid18}. Here we explain why
we came to the opposite conclusion and argue that the results reported in \cite%
{Max18} must be reconsidered by removing \textquotedblleft
corrupted\textquotedblright\ frequency intervals from the LIGO-Livingston
strain time series.

For the GW170817 event involving low mass neutron stars the gravitational
waveform is known analytically for almost the entire time interval when the
signal passes through the detector's sensitivity band (apart from the last
second before merger). This is the case because the energy loss by the
binary system is accurately described by the quadrupole formula upto the
last second. Both in general relativity and vector gravity \cite{Svid17} the
waveform is given by%
\begin{equation}
s(t)\propto \frac{1}{\left( t_{c}-t\right) ^{1/4}}\cos \left[ 2\phi (t)%
\right] ,  \label{t1}
\end{equation}%
where 
\begin{equation}
\phi (t)=-\left( \frac{c^{3}}{5GM_{c}}\right) ^{5/8}\left( t_{c}-t\right)
^{5/8}+\phi _{0},  \label{t2}
\end{equation}%
$t_{c}$ is the coalescence time and $M_{c}$ is the chirp mass of the binary
system.

We use published strain time series for the three detectors \cite{LV17}
which are not normalized to the detector's noise. The best fit to the
LIGO-Livingston and LIGO-Hanford data yields $M_{c}=1.188$ $M_{\odot }$ and $%
t_{c}=0.296$ s (here we made an adjustment for different arrival times of
the signal at the detector locations).

Next we introduce an integrated complex interferometer response 
\begin{equation}
I(t_{0},\Delta t)=\int_{t_{0}}^{t_{0}+\Delta t}\left( t_{c}-t\right)
^{1/4}e^{-2i\phi (t)}h(t)dt,  \label{x3}
\end{equation}%
where $\Delta t$ is the signal collection time and $h(t)$ is the strain
measured by the interferometer that contains both signal and noise.
According to Eqs. (\ref{t1}), (\ref{t2}) and (\ref{x3}), the signal
contribution to $I$ is proportional to $\Delta t$ provided we disregard a
small correction produced by the fast oscillating term. Thus, the signal
accumulates with an increase of the collection time $\Delta t$. In contrast,
noise does not accumulate with $\Delta t$ and for large enough $\Delta t$
the noise contribution to $I$ can be disregarded.

Thus, the theory predicts that the ratio 
\begin{equation}
u=\frac{I(t_{0},\Delta t)}{\Delta t}
\end{equation}%
should be independent of the signal collection time interval $%
[t_{0},t_{0}+\Delta t]$. This ratio can be interpreted as a signal per unit
time. It depends, in particular, on GW polarization which allows us to
distinguish between pure tensor and pure vector polarizations using the
three interferometer network \cite{Svid18a}. In addition, $u$ should be used
to determine how much signal is present in the interferometer data stream at
different times. If noise filtering has not altered the signal at certain
frequencies then $u$ must have the same (complex) value for any collection
time interval.

\begin{figure}[h]
\begin{center}
\epsfig{figure=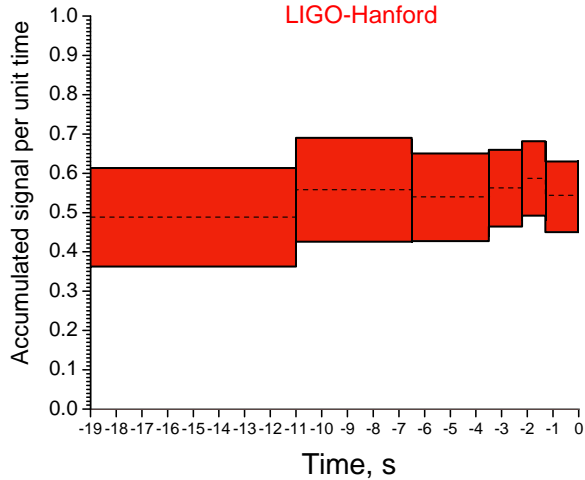, angle=270, width=9.5cm}
\end{center}
\par
\vspace{-0.8cm}
\caption{Absolute value of the signal per unit time $|u|$ measured by
LIGO-Hanford detector for the GW170817 event for different collection time
intervals. As in Ref. \protect\cite{Abbo17b}, times are shown relative to
August 17, 2017 12:41:04 UTC.}
\label{CFig1}
\end{figure}

\begin{figure}[h]
\begin{center}
\epsfig{figure=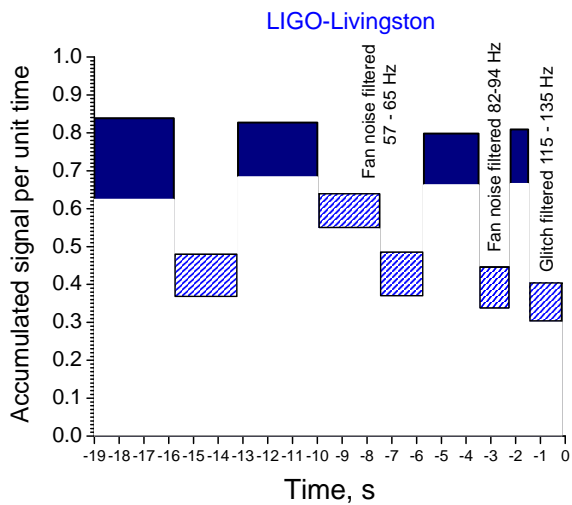, angle=270, width=9.5cm}
\end{center}
\par
\vspace{-0.8cm}
\caption{ The same as in Fig. \protect\ref{CFig1}, but for the
LIGO-Livingston detector. The scale of the vertical axis is arbitrary but
the same in both figures. Frequency regions in which signal is reduced by
noise filtering are indicated as dashed bars.}
\label{CFig2}
\end{figure}

In Fig. \ref{CFig1} we plot $|u|$ for LIGO-Hanford detector for different
collection time intervals. The result is shown as a set of rectangular bars.
The length of a bar is equal to the collection time $\Delta t$, while the
height corresponds to the uncertainty produced by the detector noise.
Namely, the half-height is equal to one standard deviation. We calculated
the error by injecting a test signal into the measured strain time series.
The average value of $|u|$ for each interval $\Delta t$ is indicated by a
dashed line. The figure shows that $|u|$ for the LIGO-Hanford detector is
consistent with a constant for the entire time when the signal is present in
the data stream. Therefore, the signal in the LIGO-Hanford strain time
series appears not to be affected by the noise subtraction.

In Fig. \ref{CFig2} we plot $|u|$ for LIGO-Livingston detector. The error
bars for this detector are smaller due to lower noise. Solid bars indicate
regions for which signal is clearly visible in the LIGO-Livingston
spectrogram \cite{Svid18a}. The vertical position of the solid bars is
consistent with $u$ being a constant, as predicted by the theory. Dashed
bars correspond to gap regions in the spectrogram \cite{Svid18a}. Fig. \ref%
{CFig2} shows that signal in these regions is substantially smaller than the
signal content in other intervals.

The signal reduction can be attributed to noise filtering. For example, a
glitch which occurred in LIGO-Livingston detector about $1.1$ s before
coalescence can explain why there is less signal in the data stream during
the glitch duration. Namely, the glitch removal from the data stream led to
some of the GW signal being subtracted off unintentionally, as one can see
from Fig. \ref{CFig2}. Signal reduction in other frequency regions of the
LIGO-Livingston detector can be attributed to filtering fan noise (see Fig.
3 in \cite{Cost18}).

The GW170817 event allows us to determine the amount by which GW signal is
suppressed by the noise filtering. Perhaps noise filtering yields
substantial signal reduction at certain frequencies for GW events for which
signal per unit time is very weak, which is the case for GW170817. Thus, for
such events one should perform a consistency check of Figs. \ref{CFig1} and %
\ref{CFig2}, and remove \textquotedblleft corrupted\textquotedblright\
frequency intervals from the data analysis. This must be taken into account
in the analysis of future GW events produced by inspiral of low-mass objects.

Contrary to the amplitude situation, we found that signal phase was not
altered in both detectors, namely the phase of $u$ is consistent with a
constant. This explains why sky location of the source is predicted
correctly from the measured strain time series. The sky localization is
constrained primarily by the differences in arrival times of the signal in
various detectors \cite{Fair18}. For the GW170817 event the time delays are
obtained with a high accuracy by fitting the signal phase $\phi (t)$ for
hundreds GW cycles. More importantly, the GW source location is known to
high accuracy thanks to the observation of concomitant electromagnetic
emission.

As we showed in \cite{Svid18a}, the ratio of LIGO-Hanford and
LIGO-Livingston signal amplitudes $|H/L|$ is crucial for distinguishing
between tensor and vector GW polarizations. In Ref. \cite{Svid18a} we
obtained this ratio by accumulating the LIGO-Livingston signal only from the
regions shown by the solid bars in Fig. \ref{CFig2}. In these regions the
signal is not corrupted. We found $0.6<\left\vert H/L\right\vert <0.8$. This
estimate, combined with other constraints, is compatible with vector theory
of gravity \cite{Svid17,Svid18} but rules out general relativity \cite%
{Svid18a}.

However, signal accumulation from the entire frequency band erroneously
underestimates the LIGO-Livingston signal amplitude yielding $1<\left\vert
H/L\right\vert <1.2$ \cite{Remark}. This incorrect estimate results in the
opposite conclusion about GW polarization. Namely, it rules out vector
polarization, but is consistent with tensor polarization (see Fig. 17 in 
\cite{Svid18a}). This is what the authors of Ref. \cite{Max18} have found.
We believe that if the LIGO-Virgo analysis were done taking into account the
Livingston signal amplitude reduction in certain frequency regions, they
would obtain the opposite result for the GW polarization, which would agree
with our findings \cite{Svid18a}.


\begin{thebibliography}{9}
\bibitem{Remark} To obtain this estimate we applied data whitening procedure
which assigns larger weight to frequency regions with better detector strain
sensitivity.
\end{thebibliography}
\end{document}